\documentclass[twocolumn,showpacs,prb,amsmath,amssymb,longbibliography]{revtex4-2}
\usepackage{graphicx}% Include figure files
\usepackage[utf8]{inputenc}
\usepackage{amsmath}
\usepackage{array}

\begin{document}

\title{Electronic structure of InP/ZnSe quantum dots: effect of tetrahedral shape, valence band coupling and excitonic interactions}

\author{Josep Planelles}
\affiliation{Dept. de Qu\'imica F\'isica i Anal\'itica, Universitat Jaume I, 12080, Castell\'o, Spain}

\author{Juan I. Climente}
\affiliation{Dept. de Qu\'imica F\'isica i Anal\'itica, Universitat Jaume I, 12080, Castell\'o, Spain}
\email{climente@uji.es}

\date{\today}

\begin{abstract}
The energy levels and optical transitions of tetrahedral core/shell InP/ZnSe quantum dots (QDs) are investigated by means of multi-band k$\cdot$p theory.
 Despite the $\overline{T}_d$ symmetry relaxing spherical selection rules, the near-band-edge excitonic spectrum is reminiscent of that obtained for spherical nanocrystals. 
Exceptions appear in large (red-emitting) QDs, where transitions violating the (quasi-)angular momentum selection rule ($\Delta L=0,\pm 2$) 
are observed, and the ground state does not become dark ($P_{3/2}$-like).
 Valence band coupling is important in determining the symmetry, degeneracy and energy of hole states,
 with split-off holes playing a greater role than in CdSe QDs.
 The ($1S_e$-like) electron ground state exhibits moderate delocalization into the ZnSe shell.
  The confinement regime is then strong even for thick shells,
 which results in Coulomb interactions being mostly perturbative.
 Electrons remain largely localized in the InP core even in negative trions, despite electron-electron repulsions.
 At the same time, the asymmetry between Coulomb attractions and repulsions leads to 
 negative (positive) trions being bound (antibound) by tens of meV. 
The biexciton binding energy switches from positive to negative, depending on the QD size.

\end{abstract}

\maketitle

\section{Introduction}

InP/ZnSe nanocrystals offer a compelling alternative to traditional cadmium-based QDs due to their lower toxicity and comparable optoelectronic properties.\cite{AlmeidaNR,LiuAOM}
The core/shell architecture of InP/ZnSe QDs, where an InP core is encapsulated by a ZnSe shell, effectively passivates surface defects and enhances photoluminescence quantum yield, making them highly suitable for applications in light-emitting diodes, bioimaging and photovoltaics.\cite{AlmeidaNR,LiuAOM,WonNAT,ReidNL,YuJPCL,LiJACS,YuLSA} The synthesis of these QDs typically involves colloidal methods, which allow for precise control over the core and shell sizes, as well as on their interface, thereby enabling tunable emission with relatively narrow linewidths across the visible spectrum.\cite{TessierCM,ToufanianFC,HahmCM,vanAvermaetACS,InadaJPCc,GiordanoCM} 

Reliable modeling of the electronic structure of InP/ZnSe QDs is a pre-requesite to rationalize and guide further experimental progress. 
To date, efforts in this direction have included effective mass (single-band, k$\cdot$p)\cite{ToufanianFC,HahmCM,JangOME} 
and density functional theory\cite{DuRSCadv} studies of hetero-nanocrystals.
These works underline the parallelism between InP/ZnSe QDs and the well-known system of CdSe/CdS QDs. 
That is, holes are largely localized inside the InP core, while electrons --having a much lower effective mass-- 
show some penetration into the ZnSe shell. %, which is reminiscent of quasi-type-II band alignment.
As a result, the ZnSe shell not only prevents traps on the InP core\cite{YuJPCL,KumarNL}, 
but it also provides a means of modulating the excitonic energy and oscillator strength. 
 Experiments are however contradictory at this regard. One the one hand, exciton emission is sensitive to the
first monolayers of ZnSe shell only\cite{ToufanianFC,vanAvermaetACS}. %, which suggests a system close to a type-I band alignment. 
On the other hand, trion and biexciton Auger rates show substantial sensitivity to the shell thickness\cite{KimNL,ReidNL,NguyenJPCc,InadaJPCc}.
%which points at a quasi-type-II band alignment. 
 The question arises whether electron-electron repulsions induce a transition from type-I to quasi-type-II behavior.\cite{KimNL}

Recently, we have employed a multi-band k$\cdot$p Hamiltonian to describe the energy levels of (non-interacting) electrons and holes in InP/ZnSe QDs.\cite{RespektaACS}
By comparing with one and two-photon absorption spectra, as well as with transient absorption measurements, 
it was established that the band offset between the core and shell materials departs from natural band offsets calculated from first principles 
for InP/ZnSe semi-bulk interfaces.\cite{wei1998calculated} 
Specifically, the valence band offset is deeper (at least $0.69$ eV), and the conduction band offset is shallower (at most $0.71$ eV).
This deviation may arise from dipoles at the InP-ZnSe interface, and/or from lattice-mismatch strain.\cite{JangOME} 
With such band offsets, a spectral assignment was proposed. 
%The band-edge absorption resonance of was identified as a $1S_{3/2}-1S_e$ transition, and the first excited one as $2S_{3/2}-1S_e$.  
 The assignment was conducted assuming spherical shape,
and hence the same symmetry labeling as for CdSe QDs.\cite{NorrisPRB,EfrosPRB}

Several questions remain however to be answered.
(i) InP and InP/ZnSe QDs tend to present tetrahedral shape.\cite{AlmeidaNR,DumbgenCM,TessierCM}
What is the impact on the energies, degeneracies and optical selection rules, as compared to the spherical approximation?
(ii) How relevant is the valence band mixing for typical QD sizes, and near band edge emission?
This aspect has been overlooked in earlier simulations,\cite{ToufanianFC,HahmCM,JangOME} 
but it was critical to understand excitonic features of Cd chalcogenide QDs.\cite{ekimov1993absorption,NorrisPRB,EfrosPRB,FonoberovPRB}
(iii) Is the spectral assignment of Ref.~\onlinecite{RespektaACS} still valid when excitonic interactions are considered?
(iv) Do carrier-carrier interactions induce a transition from type-I system (for excitons) 
towards a quasi-type-II system (for negative trions and biexcitons)? % , as suggested by some experiments\cite{KimNL}? %,NguyenJPCc}? 

In this work, we answer the questions above. To that end, we compare the energy structure and optical spectrum
of spherical and tetrahedral InP/ZnSe QDs, using a multi-band k$\cdot$p Hamiltonian and configuration interaction (CI) method
to account for carrier-carrier interactions. 

We find that, despite the different envelope symmetry and the strong confinement,
the optoelectronic properties of tetrahedral InP/ZnSe QDs are reminiscent of those of 
zinc-blende QDs with spherical shape, both at single- and multi-particle level. 
Qualitative differences arise mainly for large (dark-red emitting) QDs, 
and include the appearance of otherwise forbidden transitions, as well as the absence of a 
dark ($P_{3/2}$) ground state in the valence band. 
Valence band mixing plays an important role in determining the hole ground state energy and degeneracy,
which deviates from single-band model expectations. %and may be behind the large degeneracy observed
%in shell-filling experiments.\cite{VelosaAOM}
%%
%The dipole selection rules in the double group of the tetrahedron are more permissive than in the
%spherical approximation.
%However, a comparison between the absorption spectra of spherical and tetrahedral QDs reveals that the 
%spherical selection rules hold approximately except for the largest InP cores (dark-red emitting QDs).
%%
%The valence band structure is also reminiscent of spherical systems.
%For a single hole, heavy hole (HH) and light hole (LH) subbands couple strongly,
%with split-off holes (SOH) having significant influence too.
%%
%The ground state has $\Gamma_8$ symmetry (four-fold degenerate, akin to the $J=3/2$ angular momentum).
%%
%Contrary to expectations for spherical InP QDs, which predict a ground state transition from $S_{3/2}$ to $P_{3/2}$ 
%for large sizes \cite{EfrosPRB}, no crossing of the ground state takes place in tetrahedral QDs.
%%
%The two-hole ground state (affecting positive trions and biexcitons) is 5-fold degenerate, 
%again akin to the $J=2$ ground state of spherical CdSe QDs.\cite{RodinaPRB}
%This may explain observations in shell filling experiments of InP/ZnSe QDs, 
%which pointed at hole ground state degeneracies exceeding 4.\cite{VelosaAOM}
 %
 The inclusion of Coulomb interactions produces a fairly rigid redshift of the exciton absorption spectra,
 with scarce influence on the energy splitting between low-energy peaks. 
 This reinforces the spectral assignment we proposed in Ref.~\onlinecite{RespektaACS}.
 At the same time, repulsions give rise to large antibinding energies for positive trions and biexcitons, and to moderate binding energies for negative trions.
 Changes in the electronic density are however minor, as the quantum confinement exerted by the core remains as the main physical force.

\section{Theoretical Model}
\label{s:theo}

\subsection{Single particle states}

Non-interacting electron and hole states in zinc-blende heterostructures
are calculated using two-band ($\Gamma_6$) and six-band 
(coupled $\Gamma_8$ and $\Gamma_7$ bands) k$\cdot$p Hamiltonians.
Thus, we aim at obtaining spinorial wave functions of the form: 
\begin{align}
%	|n_e,\uparrow \rangle  &= f_{n_e}(\mathbf{r}_e) \, |u_1\rangle , \\
%	|n_e,\downarrow \rangle  &= f_{n_e}(\mathbf{r}_e) \, |u_2\rangle , \\
	|n_e \rangle  &= \sum_{i=1}^2 f_{n_e,i}(\mathbf{r}_e) \, |u_i\rangle , \\
	|n_h \rangle &= \sum_{i=3}^8 f_{n_h,i}(\mathbf{r}_h) \, |u_i\rangle .
\end{align}
\noindent Here, $f_{n_e,i}(\mathbf{r})$ are the electron envelope functions,
and $f_{n_h,i}(\mathbf{r})$ those of the hole. 
 $|u_i\rangle$ stands for the Bloch functions, defined as in Ref.\cite{NovikPRB}.
 For electrons, they have $\Gamma_6$ symmetry and Bloch angular momentum $j=1/2$:
\begin{eqnarray}
	\nonumber
	|u_1\rangle &=& |\Gamma_6,\, 1/2, +1/2\rangle = S \uparrow , \\
	|u_2\rangle &=& |\Gamma_6,\, 1/2, -1/2\rangle = S \downarrow. 
	\label{eq:elec}
\end{eqnarray}
\noindent For heavy holes (HH) and light holes (LH), they have $\Gamma_8$ symmetry and Bloch angular momentum $j=3/2$:
\begin{eqnarray}
	\nonumber
	|u_3\rangle &=& |\Gamma_8,\, 3/2, +3/2\rangle = (1/\sqrt{2})\,(X+iY)\,\uparrow, \\
	\nonumber
	|u_4\rangle &=& |\Gamma_8,\, 3/2, +1/2\rangle = (1/\sqrt{6})\,\left[(X+iY)\,\downarrow -2 Z \uparrow \right], \\
	\nonumber
	|u_5\rangle &=& |\Gamma_8,\, 3/2, -1/2\rangle = -(1/\sqrt{6})\,\left[(X-iY)\,\uparrow +2 Z \downarrow \right], \\
	|u_6\rangle &=& |\Gamma_8,\, 3/2, -3/2\rangle = -(1/\sqrt{2})\,(X-iY)\,\downarrow. 
	\label{eq:HHLH}
\end{eqnarray}
\noindent For split-off holes (SOH), they have $\Gamma_7$ symmetry and Bloch angular momentum $j=1/2$:
\begin{eqnarray}
	\nonumber
	|u_7\rangle &=& |\Gamma_7,\, 1/2, +1/2\rangle = (1/\sqrt{3})\,\left[(X+iY)\,\downarrow + Z \uparrow \right], \\
	|u_8\rangle &=& |\Gamma_7,\, 1/2, -1/2\rangle = (1/\sqrt{3})\,\left[(X-iY)\,\uparrow - Z \downarrow \right].
	\label{eq:SOH}
\end{eqnarray}

The corresponding hamiltonian for electrons (two-band but diagonal) is:
\begin{equation}\label{eq:He}
	H_e = \frac{\hbar^2}{2\,m_0}\,\left( \mathbf{p} \, \frac{1}{m_e(\mathbf{r})} \, \mathbf{p} \right) + V_e(\mathbf{r}),
\end{equation}
\noindent with $m_0$ the free-electron mass, $\mathbf{p}$ the momentum operator, $m_e(\mathbf{r})$ the position-dependent 
effective mass of the electron and $V_e(\mathbf{r})$ the potential felt by the electron: 
\begin{equation}\label{eq:pot}
	V_e(\mathbf{r})= \left\lbrace 
	\begin{array}{ll}
		E_{\text{g}}^{\text{InP}} & \text{if $\mathbf{r} \in$ InP,} \\
		E_{\text{g}}^{\text{InP}}+cbo & \text{if $\mathbf{r} \in$ ZnSe,} 
	\end{array}
	\right.
\end{equation}
\noindent where $E_{\text{g}}^{\text{InP}}$ is the bulk band gap of InP, and $cbo$ the InP/ZnSe conduction band offset.

The hole Hamiltonian, projected onto $\{|u_3\rangle,\ldots,|u_8\rangle\}$,
with $z$ along the $[001]$ crystallographic direction, reads:\cite{NovikPRB}
\begin{equation}\label{eq:Hh}
	H_h = 
	\left(
\small
	\begin{array}{cccccc}
		U+V   &   -S     &   R    &    0    & \frac{1}{\sqrt{2}}\,S   & -\sqrt{2} R \\
	  -S^\dagger  &    U-V   &   0    &    R    &    \sqrt{2}\,V          & -\sqrt{\frac{3}{2}}\,S  \\
	R^\dagger     & 0        &  U-V   &    S    &   -\sqrt{\frac{3}{2}}\,S^\dagger & -\sqrt{2}\,V \\
		0            &   R^\dagger    & S^\dagger &  U+V  & \sqrt{2}\,R^\dagger & \frac{1}{\sqrt{2}}\,S^\dagger \\
		\frac{1}{\sqrt{2}}\,S^\dagger &  \sqrt{2}\,V & -\sqrt{\frac{3}{2}}\,S & \sqrt{2}\,R & U-\Delta & 0 \\
		-\sqrt{2}\,R^\dagger & - \sqrt{\frac{3}{2}}\,S^\dagger & -\sqrt{2}\,V & \frac{1}{\sqrt{2}}\, S & 0 & U-\Delta
	\end{array}
\normalsize
	\right).
\end{equation}
\noindent Here, $\Delta$ is the spin-orbit energy separating SOH from HH and LH subbands. 
The other terms are: 
\begin{align}
	\label{eq:U}
	U = & -\frac{\hbar^2}{2m_0}\, \gamma_1 p^2 + V_h(\mathbf{r}), \\
	\label{eq:V}
	V = & -\frac{\hbar^2}{2m_0}\, \gamma_2 (p_\parallel^2 -2 p_z^2 ), \\
	\label{eq:R}
	R = & -\frac{\hbar^2}{2m_0}\, \sqrt{3}\,\left( \mu\,p_+^2 - \bar{\gamma}\,p_-^2 \right), \\
	\label{eq:S}
    	S = & -\frac{\hbar^2}{ m_0}\, \sqrt{3}\,\gamma_3 p_- p_z,
\end{align}
\noindent with $\gamma_1$,$\gamma_2$,$\gamma_3$ the Luttinger parameters,
$\mu = (\gamma_3 - \gamma_2)/2$ and $\bar{\gamma}=(\gamma_3+\gamma_2)/2$.
$p^2=p_x^2+p_y^2+p_z^2$, $p_\parallel=p_x^2+p_y^2$ are the components of the
squared momentum and $V_h(\mathbf{r})$ the hole confining potential:
\begin{equation}\label{eq:poth}
	V_h(\mathbf{r})= \left\lbrace 
	\begin{array}{ll}
		0  & \text{if $\mathbf{r} \in$ InP,} \\
		vbo & \text{if $\mathbf{r} \in$ ZnSe,} 
	\end{array}
	\right.
\end{equation}
\noindent where $vbo$ is the bulk valence band offset between InP and ZnSe.
Because $vbo$ is quite deep, holes are expected to be largely confined inside the InP core.
For this reason, we have simplified $H_h$ by considering homogeneous (InP) 
Luttinger parameters, rather than position-dependent ones. 
Lattice-mismatch strain and self-energy terms are not accounted for explicitly. 
Their main impact is to lower $cbo$.\cite{JangOME,FonoberovPRB} 
Because we infer this parameter from comparison with experiments,\cite{RespektaACS}
 the effect is implicit.

 Hamiltonians $H_e$ and $H_h$ are integrated numerically, using the finite elements method
 with Comsol Multiphysics.

\subsection{Many-body states}

The Hamiltonian for excitons ($X$), biexcitons ($XX$), positive trions ($X^+$) and negative trions ($X^-$)
can be written in the second quantization as:

\begin{multline}
{\hat H} =  \sum_i E_i^e\,e_i^+\,e_i  +  \sum_i E_i^h\,h_i^+\,h_i  \\
+ \sum_{ijkl} \langle i|\langle j| \,W_{eh}\, | k\rangle |l\rangle \, e_i^+  h_j^+  h_k  e_l \\
+ \frac{1}{2} \sum_{ijkl} \langle ij| \,W_{ee}\, | kl\rangle \, e_i^+  e_j^+  e_k  e_l \\
+ \frac{1}{2} \sum_{ijkl} \langle ij| \,W_{hh}\, | kl\rangle \, h_i^+  h_j^+  h_k  h_l ,
\label{Hfull}
\end{multline}

\noindent where $E_i^e$ ($E_i^h$) is the electron (hole) energy in the
single-particle state $|i\rangle$, $e_i$ ($h_i$) is the electron
(hole) annihilation operator and $\langle ij|\,W\,| kl \rangle$ is the
two-body Coulomb matrix element. Electron-hole exchange terms are neglected.
Coulomb matrix elements are calculated by integrating Poisson equation
in a heterogeneous dielectric environment. Specifically, we consider
a different dielectric constant in the inorganic QD ($\varepsilon_{in}$)
and the surrounding --usually organic-- medium ($\varepsilon_{out}$). 
Because $\varepsilon_{out} \ll \varepsilon_{in}$, this dielectric mismatch 
greatly enhances Coulomb interactions.\cite{FonoberovPRB,RodinaJETP}
 Hamiltonian (\ref{Hfull}) is solved using a CI method.\cite{CItool}

 \subsection{Optical transitions and selection rules}\label{ss:rules}

 The probability of a transition from an initial state $|i\rangle$ to a final state $|f\rangle$
 is calculated using Fermi's golden rule:
 \begin{equation}
	 P_{i \rightarrow f} (h \nu) \propto \sum_\alpha \left| \langle f | \epsilon_\alpha p_\alpha | i \rangle \right|^2 \, \delta(E_{if} -h \nu) \, n_i(T)\,(1-n_f(T)).
 \end{equation}
 \noindent Here,
 $\alpha$ runs over the different (orthogonal) components of the polarization vector $\boldsymbol{\epsilon}$.
% ($\epsilon_+ = 1/\sqrt{2}\,(1,i,0)$, $\epsilon_- = 1/\sqrt{2}\,(1,-i,0)$, $\epsilon_z = (0,0,1)$),
%
 $n_j$ is the thermal population of the state $|j\rangle$ at temperature $T$,
 and $E_{if}=E_f-E_i$ is the energy difference between states $|i\rangle$ and $|f\rangle$.
 In the simulations, the resonance condition, $\delta(E_{if}-h\nu)$, is replaced by a Gaussian function
 of bandwidth 1 meV.
 The interband absorption spectrum is obtained computing $P_{i \rightarrow f}$ 
 for all the possible transitions between near-band-edge initial excitonic states with $N_e$ electrons
 and $N_h$ holes, and final states with $N_e-1$ electrons and $N_h-1$ holes.

 Selection rules arise from the dipole matrix element, $\langle f | \mathbf{p} | i \rangle$. 
 In InP/ZnSe QDs, both the microscopic lattice\cite{ReidNL} and the
 InP core shape\cite{AlmeidaNR,DumbgenCM,TessierCM,KumarNL} have tetrahedral symmetry.
 Thus, the relevant point group is the double group of the tetrahedron,  $\overline{T}_d$. 
 Transitions are dipole-allowed if the integrand in $\langle f | \mathbf{p} | i \rangle$
 belongs to the totally symmetric irreducible representation, $\Gamma_1$.
 %
% The momentum operator $\mathbf{p}$ has $\Gamma_5$ symmetry ($T_2$ in Schoenflies notation).

 Often, more specific selection rules can be derived by integrating envelope and Bloch parts separately.
 For example, an interband transition in the independent particle scheme is given by
 the coupling between a valence band hole state $|n_h\rangle$ and a conduction band electron state $|n_e\rangle$:\cite{bastard1990wave}
\begin{equation}
	\langle n_e | p_\alpha | n_h \rangle  = 
	\sum_{j=1}^2 \sum_{i=3}^8  \langle f_{n_e,j} | f_{n_h,i} \rangle \, \langle u_j | p_\alpha | u_i \rangle .
\end{equation}
The Bloch integral, $\langle u_j | p_\alpha | u_i \rangle$, accounts for the
selection rules arising from the lattice symmetry. They can be derived from 
Eqs.~(\ref{eq:elec})-(\ref{eq:SOH}). In the $\overline{T}_d$ group, the 
momentum operator has $\Gamma_5$ symmetry.
Because the products $\Gamma_6 \times \Gamma_5 \times \Gamma_8$ and
 $\Gamma_6 \times \Gamma_5 \times \Gamma_7$ contain the totally symmetric representation
(see Table \ref{tab:prods} in Appendix \ref{ap:prods}), transitions to the conduction band are allowed from 
HH, LH and SOH bulk bands alike. 

\begin{table}[b]
\caption{Envelope, Bloch and total irreducible representations of the lowest conduction ($\Gamma_6$), HH-LH ($\Gamma_8$) and SOH ($\Gamma_7$) states
	in $\overline{T}_d$. The envelope angular momentum $L$ in the spherical limit is also given. }\label{tab:irreps}
\begin{center}
	\begin{tabular}{|p{0.75cm}|l|l|l|}
\hline
\centering $L$ & Envelope & Bloch  & Total  \\
\hline
\hline
\centering 0 & $\Gamma_1$ & $\Gamma_6$ & $\Gamma_6$ \\
\hline
\centering 1 & $\Gamma_5$ & $\Gamma_6$ & $\Gamma_7 \oplus \Gamma_8$ \\
\hline
\centering 2 & $\Gamma_3 \oplus \Gamma_5$ & $\Gamma_6$ & $\Gamma_7\oplus 2\Gamma_8$ \\
\hline
\hline
\centering 0 & $\Gamma_1$ & $\Gamma_8$ & $\Gamma_8$ \\
\hline
\centering 1 & $\Gamma_5$ & $\Gamma_8$ & $ \Gamma_6 \oplus  \Gamma_7 \oplus 2\Gamma_8$ \\
\hline
\centering 2 & $\Gamma_3 \oplus \Gamma_5$ & $\Gamma_8$ & $2 \Gamma_6 \oplus 2 \Gamma_7 \oplus 3\Gamma_8$ \\
\hline
\hline
\centering 0 & $\Gamma_1$ & $\Gamma_7$ & $\Gamma_7$ \\
\hline
\centering 1 & $\Gamma_5$ & $\Gamma_7$ & $\Gamma_6 \oplus \Gamma_8$ \\
\hline
\centering 2 & $\Gamma_3 \oplus \Gamma_5$ & $\Gamma_7$ & $\Gamma_6 \oplus 2\Gamma_8$ \\
\hline
\end{tabular}
\end{center}
\end{table}
 
The envelope integral, $\langle f_{n_e,j} | f_{n_h,i} \rangle$ provides an additional restriction,
related to the QD confinement. Namely, both electron and hole envelope functions must have the same symmetry.
 When cubic band warping is neglected, $\mu=0$ in Eq.~(\ref{eq:R}), and the confining potential is spherical, 
 $H_e$ and $H_h$ present spherical symmetry.
 Electron states can then be labeled as $n_e\,L_e$, where $L_e$ is the envelope angular momentum.
 The lowest states are typically $1S_e$, $1P_e$, $2S_e$, etc.
 Hole states, under the influence of band mixing, have well defined total angular momentum $J_h=L_h+j_h$, 
 with $L_h$ the angular momentum and $j_h$ the Bloch one. The states can be labeled as
 $n_h\,(L_h)_{J_h}$, where $L_h$ is the lowest angular momentum of the spinor.
 Top-most states include $1S_{3/2}$, $1P_{3/2}$, $1S_{1/2}$ and $1P_{1/2}$.\cite{EfrosPRB} %It is important to note that the envelope functions of these states
 Within these states, valence band mixing couples envelope components differing in $\Delta L_h=0,\pm 2$.\cite{BaldereschiPRB}
% Thus, different components $f_{n_h,i}$ of a given state $|n_h\rangle$ have different $L_h$.
 %
 It follows that $\langle f_{n_e,j} | f_{n_h,i} \rangle \propto \delta_{L_e,L_h}$,
 and allowed transitions include $n_h\,L_{J_h} \rightarrow n_e L$ and 
  $n_h\,L_{J_h} \rightarrow n_e (L+2)$.
 For example, $1S_{3/2}$ has a HH component with $L_h=0$ and a LH one with $L_h=2$.
 Then, $1S_{3/2} 1S_e$ and $1S_{3/2} 1D_e$ are both allowed.
 These strict selection rules have provided a successful assignment of the absorption spectra of CdSe QDs,\cite{ekimov1993absorption,NorrisPRB}
 and were recently used for InP/ZnSe QDs\cite{RespektaACS}. 

 In InP QDs, however, the confining potential is tetrahedral and band warping may be non-neglegible.  
 Because the point group is now finite, the selection rule associated with $\langle f_{n_e,j} | f_{n_h,i} \rangle$ 
 is relaxed.
 As shown in Table \ref{tab:irreps},
 former $L=0,\,L=1$ and $L=2$ orbitals acquire $\Gamma_1, \Gamma_5$ and $\Gamma_3 \oplus \Gamma_5$ symmetry, respectively.
 Consequently, transitions between $1S_{3/2}$ and $1P_e$ states, forbidden in a spherical system, become active
 under a cubic perturbation, as both involve the $\Gamma_5$ representation.
% States with $L>2$ will display increasing optical activity. 
 %
 In addition, $J$ is no longer a valid quantum number, which lifts the $\Delta L_h=0,\pm 2$ restriction
 within the envelope components of $|n_h\rangle$.

 The only general selection rule remaining is that associated with the total symmetry of 
 $\langle n_e | p_\alpha | n_h \rangle$. Table \ref{tab:irreps} shows that 
 $|n_e\rangle$ and $|n_h\rangle$ states can have total symmetries $\Gamma_6$, $\Gamma_7$ and $\Gamma_8$.
 Because the momentum operator has $\Gamma_5$ symmetry, the products $\Gamma_f \times \Gamma_5 \times \Gamma_i$
 contain $\Gamma_1$ in all instances except when $\Gamma_f = \Gamma_i = \Gamma_6, \Gamma_7$.
 Thus, transitions to $1S_e$-like states (total symmetry $\Gamma_6$) 
 are possible from any HH, LH and SOH state except those with $\Gamma_6$ symmetry.
 Transitions to $1P_e$-like states (total symmetry $\Gamma_7$) are possible
 from any hole state, except those with $\Gamma_7$ symmetry.
 Transitions to $1D_e$-like states (total symmetry $\Gamma_8$) are possible
 from any hole state, with no restriction.

\section{Results}

We study the electronic structure of electrons, holes, excitons, trions and biexcitons in InP/ZnSe QDs.
To model the InP core, we take $m_e=0.08$, $\gamma_1=5.08$, $\gamma_2=1.60$, $\gamma_3=2.1$, $\Delta=0.108$ eV 
and $E_{\text{g}}^{\text{InP}}=1.42$ eV.\cite{Vurgaftman2001band}
In the ZnSe shell, we take $m_e=0.16$ and $E_{\text{g}}^{\text{ZnSe}}=2.82$ eV,\cite{LB:ZnSe_gap,LB:ZnSe_me} 
but keep InP parameters for the valence band ($\gamma_1,\gamma_2,\gamma_3,\Delta$).
This is a reasonable approximation because holes stay largely confined in the core,
and simplifies $H_h$ by avoiding the need of Burt-Foreman boundary conditions.\cite{BurtJPCM,ForemanPRB}
 The band offsets for the InP/ZnSe interface are $vbo=0.9$ eV and $cbo=0.5$ eV, which fall within the range proposed in Ref.~\onlinecite{RespektaACS}. Relative dielectric constants are $\varepsilon_{in}=10$ for the QD and $\varepsilon_{out}=2$ for the organic environment. In all systems, we calculate the 10 lowest electron (30 highest hole) spinorial, non-interacting states. 
 For excitonic systems, CI simulations use a basis built from all Hartree products of single-particle wave functions (excitons),
 single-particle wave functions combined with Slater determinants (trions), and producs of Slater determinants (biecitons),
 all derived from the non-interacting states.\cite{CItool}

 \subsection{Non-interacting electron and hole}

\begin{figure}[h]
    \centering
    \includegraphics[width=7.5cm]{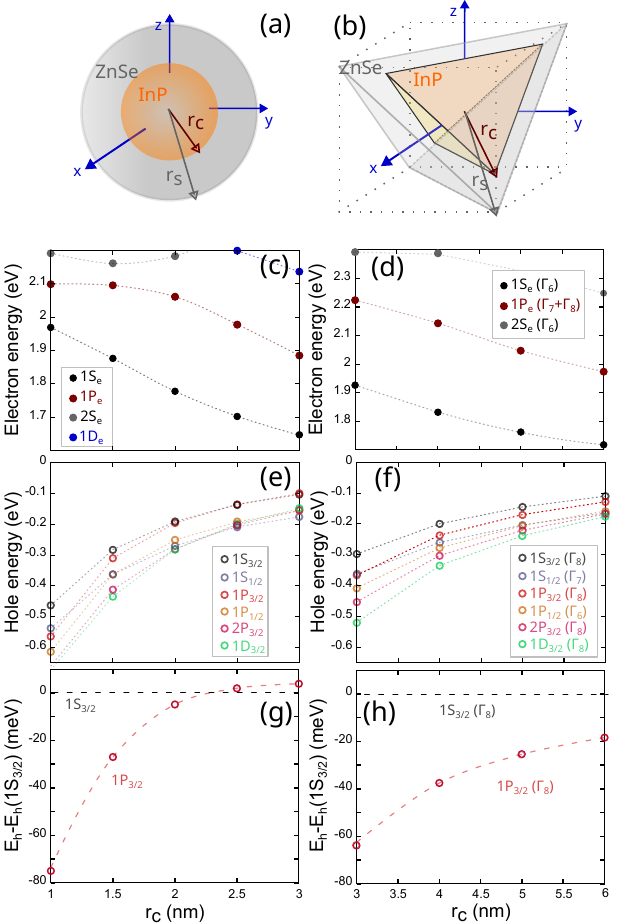}
	\caption{Electronic structure of non-interacting electron and hole.
	(a) and (b): schematic of the spherical and tetrahedral core/shell QDs under study.  
%	In the tetrahedral structure, $r_c$ and $r_s$ are distances from the center to a vertex.
	(c) and (d): electron energy levels for varying $r_c$.
	(e) and (f): hole energy levels for varying $r_c$. 
	(g) and (h): energy difference between $1P_{3/2}$ and $1S_{3/2}$ states. 
	In spherical QDs, $r_s=5$ nm. In tetrahedral QDs, $r_s=8$ nm.
	}
    \label{fig1}
\end{figure}

 We start by comparing the (near-band-edge) energy levels of non-interacting carriers in spherical (Fig.~\ref{fig1}a) 
 and tetrahedral (Fig.~\ref{fig1}b) QDs. The shell size ($r_s$) is fixed and the core size ($r_c$) is varied.
 In tetrahedral QDs, $r_c$ and $r_s$ are distances from the center to a vertex. 
 We consider dimensions such that electron-hole energies lie in the cyan blue-to-red spectral window. 
 This implies larger $r_c$ values for tetrahedra than for spheres, as their shape makes them more sensitive to confinement.

 Figs.~\ref{fig1}c and \ref{fig1}d show that electron levels display stabilization when confinement decreases, 
 but it deviates from a simple $1/r_c^2$ trend. Also, the splitting between energy levels does not decrease
 with increasing $r_c$.  These are signatures of the ZnSe shell playing a role. 
 We come back to this point below.  At this stage, we wish to underline the parallelism between
 states in spherical and tetrahedral QDs. In spherical QDs, the lowest states are easily identified as $1S_e$, $1P_e$ and $2S_e$, 
 with degeneracies of 2, 6 and 2 -- including spins.
 In tetrahedral QDs, the symmetries can be retrieved from the wave functions following the procedure
 described in Appendix \ref{ap:sim}.
 The lowest state has $\Gamma_6$ symmetry 
 (2-fold degenerate, see the dimension of the irreducible representations in the character table of $\overline{T}_d$,
 Table \ref{tab:chars} in Appendix \ref{ap:prods}), 
 then two states of nearly identical
 energy with $\Gamma_7$ (2-fold degenerate) and $\Gamma_8$ (4-fold degenerate) symmetries, and next
 another $\Gamma_6$ state. It is clear that a correspondence with the $1S_e$, $1P_e$ and $2S_e$ states
 holds despite the symmetry reduction.

 Figs.~\ref{fig1}c and \ref{fig1}d compare the hole states in both systems. 
 In spherical QDs, the top-most states are $1S_{3/2}$, $1P_{3/2}$ and $1S_{1/2}$, 
 which present the same degeneracies as their tetrahedral counterparts, with
 $\Gamma_8$, $\Gamma_8$ and $\Gamma_7$ symmetries.
 That is, a clear correspondence is found again between the spherical states and their 
 tetrahedral counterparts.
 For this reason, in what follows we label the states of both systems using the
 usual spherical notation, $n L_J$.

 A few qualitative aspects differ between spherical and tetrahedral QDs, however. 
 For example, in spherical QDs, the hole ground state switches from $1S_{3/2}$ (bright)
 to $1P_{3/2}$ (dark) when $r_c > 2$ nm.
 This is better seen in Fig.~\ref{fig1}g, which represents the energy difference with respect
 to $1S_{3/2}$.
 The switch vanishes in tetrahedral QDs, Fig.~\ref{fig1}h. 
 The reason is that $1S_{3/2}$ and $1P_{3/2}$ share $\Gamma_8$ symmetry, so that an anticrossing
 (rather than a crossing) takes place. 
 This result anticipates that a gradual change in the ground state properties of
 InP/ZnSe QDs should be expected for large enough sizes. 
 A similar effect was reported in tetrahedral CdS QDs, albeit the crossover
 there occured for decreasing (rather than increasing) radii.\cite{FonoberovPRB}

\begin{figure}[h]
    \centering
    \includegraphics[width=6.5cm]{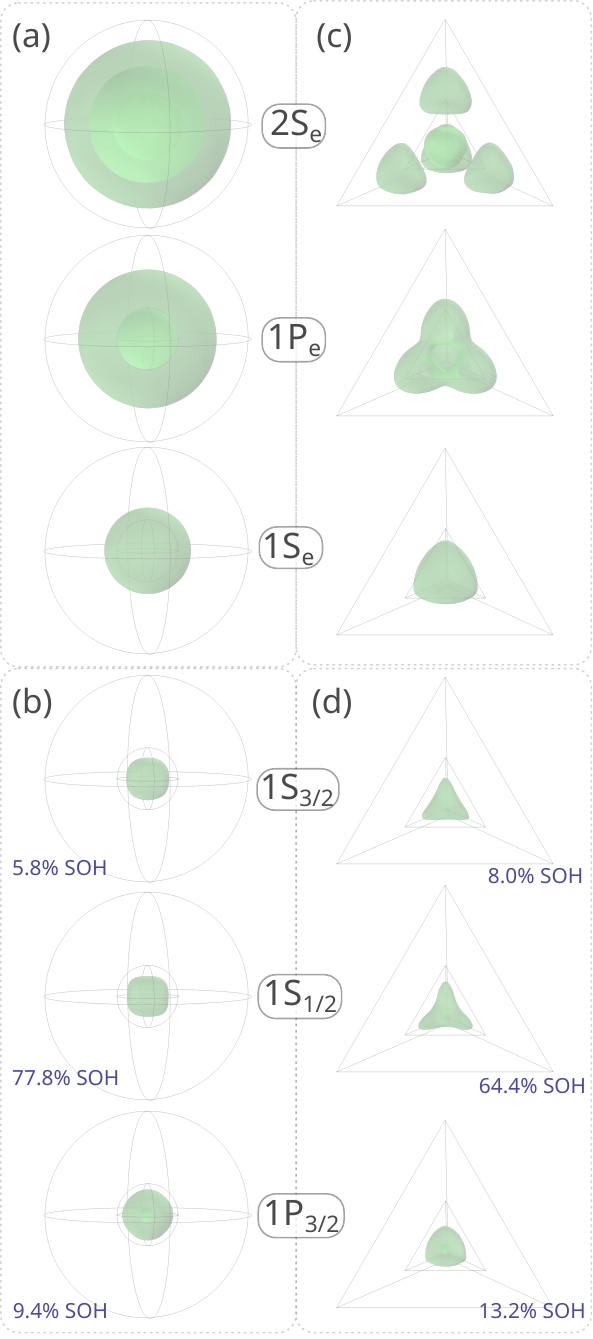}
	\caption{Charge density of near band edge electrons (a) and holes (b) in a spherical QD with $(r_c,r_s)=(1.5,5)$ nm. 
	(c) and (d): same but for a tetrahedral QD with $(r_c,r_s)=(3.0,8.0)$ nm.
	The isosurfaces contain 70\% of the charge density.}
    \label{fig2}
\end{figure}

 Insight into the role of the ZnSe shell is given in Figure \ref{fig2}.
 Figs.~\ref{fig2}a and \ref{fig2}b show charge densities of electron and hole states
 in a spherical QD with $r_c$=1.5 nm (orange-to-red emission).
 Holes are largely confined within the InP core, because $vbo$ is high and
 hole masses are heavy. By contrast, electron states penetrate into the ZnSe shell. 
 The effect is already visible for $1S_e$, but it becomes much more pronounced 
 for $1P_e$ and $2S_e$, as their energy largely exceeds $cbo$.
 The same behavior is observed in tetrahedral QDs, Figs.~\ref{fig2}c and \ref{fig2}d,
 albeit the penetration of $1S_e$ is comparatively smaller.
 The fact that $1S_e$ is largely localized in the core, while $1P_e$ and $2S_e$ are not,
 is responsible for the large energy splitting between ground and excited states in Fig.~\ref{fig1}c.
 With increasing $r_c$, $1S_e$ is stabilized but $1P_e$ and $2S_e$ less so. 
 We shall see below the energetic isolation of $1S_e$ makes Coulomb interactions perturbative. %suppresses electronic correlations.
% It follows that non-interacting electrons and holes in InP/ZnSe QDs behave 
% indeed similar to those in CdSe/CdS QDs, with a quasi-type-II band alignment.

The significance of valence band mixing can be studied by analyzing the composition of hole states. 
We find that, as in spherical CdSe QDs,\cite{EfrosPRB} HH and LH are strongly mixed, with equal contribution 
to the composition of all the states. 
Unlike in CdSe, however, SOH now has a significant influence. 
 As indicated in Fig.~\ref{fig2}b and \ref{fig2}d, the $1S_{3/2}$ ground state has sizable 
SOH character (8\% in the tetrahedral QD). %, the rest being HH-LH character. 
%As in spherical QDs,\cite{EfrosPRB} tetrahedral QDs also exhibit equal contribution of HH and LH subbands.
%
 Excited hole states present increasing SOH character, with some states like $1S_{1/2}$
 being of dominant SOH origin.
The complex valence band structure will be relevant in the spectral assignment, as we show next.

 \subsection{Excitonic spectrum}

\begin{figure}[h]
    \centering
    \includegraphics[width=7.5cm]{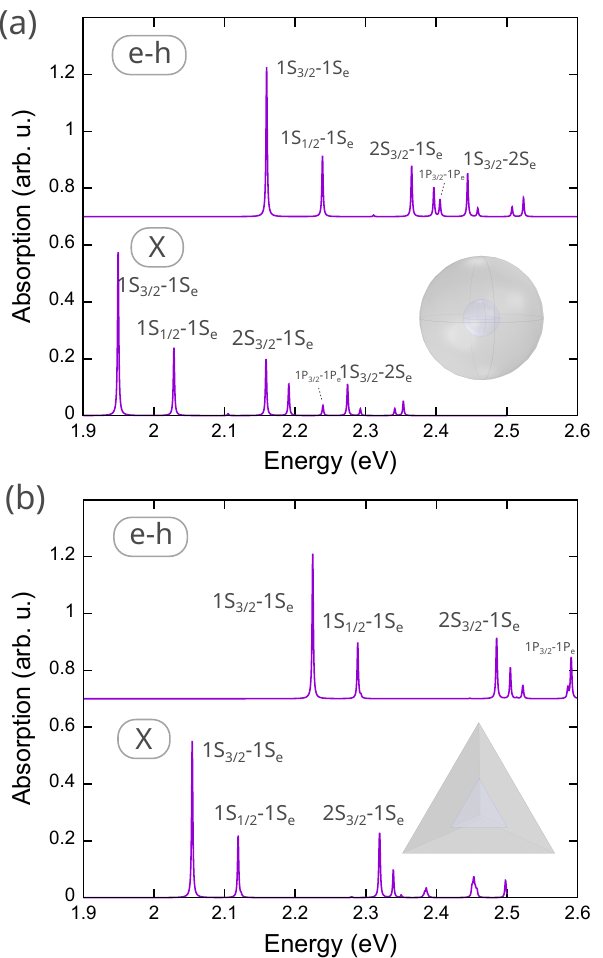}
	\caption{Spectral assignment of the lowest transitions in the absorption of InP/ZnSe QDs. 
	(a) Spherical   QD with $(r_c,r_s)=(1.5,5.0)$ nm. 
	(b) Tetrahedral QD with $(r_c,r_s)=(3.0,8.0)$ nm. 
	In each panel, the top spectrum gives the absorption of non-interacting electron-hole pairs, and the bottom one that of interacting excitons.
	The insets represent the geometry under study. 
	In the calculation, $T=0$ K.} 
    \label{fig3}
\end{figure}

In Figure \ref{fig3}, we represent the absorption spectrum of InP/ZnSe QDs,
for the same geometries studied in Fig.~\ref{fig2}. 
The top curve in Fig.~\ref{fig3}a
shows the absorption of a non-interacting electron-hole pair ($e-h$) in a spherical QD,
while the bottom curve shows that including excitonic interactions ($X$). 
Clearly, the effect of Coulomb interactions on the main transitions of the spectrum
($1S_{3/2} 1S_e$, $1S_{1/2}1S_e$, $2S_{3/2} 1S_e$, $1S_{3/2} 2S_e$) is but a fairly rigid redshift of $\sim 0.2$ eV.
In other words, excitonic interactions are strong --stimulated by dielectric confinement-- but act chiefly as a first-order perturbation. 
 This is because, as mentioned before, $1S_e$ is energetically far from the excited electron states (see Fig.~\ref{fig1}c). % ($\Delta E \geq 300$ meV, Fig.~\ref{fig1}c). 
 Then, the electron-hole attraction %($\langle V_{eh} \rangle \approx -250$ meV) 
 does not suffice to correlate $1S_e$ and $1P_e$ states.
 
When the tetrahedral shape of the QD is taken into account, Fig.~\ref{fig3}b,
the main transitions of the absorption spectrum remain the same:
$1S_{3/2} 1S_e$, $1S_{1/2} 1S_e$ and $2S_{3/2} 1S_e$.
We then conclude that the spectral assignment drafted in 
Ref.~\onlinecite{RespektaACS} is qualitatively valid in the presence of 
excitonic interactions and tetrahedral confinement.

\begin{figure}[h]
    \centering
    \includegraphics[width=7.5cm]{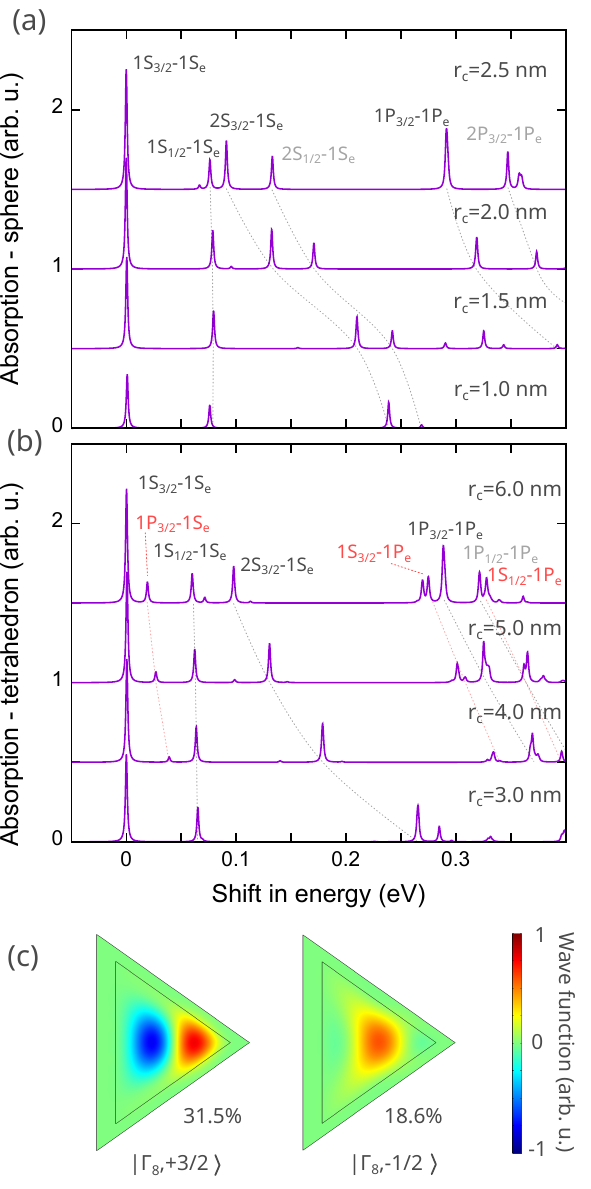}
	\caption{Absorption spectrum of $X$ in (a) spherical and (b) tetrahedral QDs
	with different core size. The spectra are offset vertically for clarity.
	Dotted lines are guides to the eyes. Red lines and characters are used
	to highlight transitions forbidden in spherical QDs.
	The energy reference is that of the $1S_{3/2} 1S_e$ transition.
	(c) Wave function of the main HH and secondary LH components of $1P_{3/2}$
	when $r_c=6$ nm. The numbers give the weight of the component within the state.}
    \label{fig4}
\end{figure}

A systematic comparison of the core size dependence in spherical and
tetrahedral QDs is given in Fig.~\ref{fig4}a and Fig.~\ref{fig4}b, respectively.
A few relevant observations can be drawn.
(i) The band edge transition ($1S_{3/2} 1S_e$, reference energy) gains
oscillator strength with increasing core size. 
This is because $1S_e$ becomes increasingly localized in the core, 
maximizing its overlap with the hole ground state.
(ii) The $1S_{1/2} 1S_e$ transition stays $60-70$ meV from the band edge
transition, irrespective of the size.
This is because $1S_{3/2}$ and $1S_{1/2}$ wave functions are similar --see Fig.~\ref{fig2}--, and hence feel confinement similarly. Their splitting is largely set by the bulk spin-orbit interaction.
(iii) By contrast, the $2S_{3/2} 1S_e$ transition approaches the band edge one with increasing $r_c$.
This is because $2S_{3/2}$, with its radial node, is more sensitive to the confinement than $1S_{3/2}$.
In the absorption experiments of Ref.~\onlinecite{RespektaACS}, 
red-emitting QDs exhibited an excited transition $0.2$ eV above the fundamental one,
whose splitting increased to $0.33$ eV in green emitting QDs.
As shown in Fig.~\ref{fig4}, both the magnitude and the size-sensitivity 
of the splitting reinforce the hypothesis that such a transition is 
$2S_{3/2} 1S_e$.
(iv) Transitions involving excited electron states (such as $1P_{3/2} 1P_e$)
are high in energy, despite the ZnSe shell relaxing the electron confinement.
These are only strong for the largest cores, as otherwise $1P_e$ delocalizes
over the shell.
(v) In tetrahedral QDs with large cores, a number of transitions build up,
which have no correspondence in spherical QDs. 
These transitions are labeled with red characters in Fig.~\ref{fig4}b.
The origin of these transitions is the relaxed selection rules 
in the $\overline{T}_d$ group, discussed in Section \ref{ss:rules}.
As mentioned above, for large cores, $1S_{3/2}$-like and $1P_{3/2}$-like states
anticross and mix (both have $\Gamma_8$ symmetry).  
Consequently, transitions such as $1P_{3/2} 1S_e$ 
($\Gamma_8 \rightarrow \Gamma_6$ in $\overline{T}_d$) become visible.
The mixing of different angular momenta in tetrahedral QDs is illustrated
in Fig.~\ref{fig4}c.
The left panel shows the wave function of the dominant HH component
in the $1P_{3/2}$ state. As expected, a $p$-like orbital can be identified. 
 The right panel, however, shows one of the LH components of the same state,
where a $s$-like function is forming.
 No such component exists when the Hamiltonian is spherical, as they 
 follow $\Delta L_h=0\,\pm 2$.\cite{BaldereschiPRB}
It is this $s$-like, totally symmetric ($\Gamma_1$) envelope component that 
enables optical coupling with the $1S_e$-like electron.
To our knowledge, no experimental evidence of these transitions 
has been reported to date. 
They likely fall within the bandwidth of more intense, 
spherical symmetry-allowed transitions,
such as the $1S_{3/2} 1S_e$ one.

\begin{figure}[h]
    \centering
    \includegraphics[width=7.5cm]{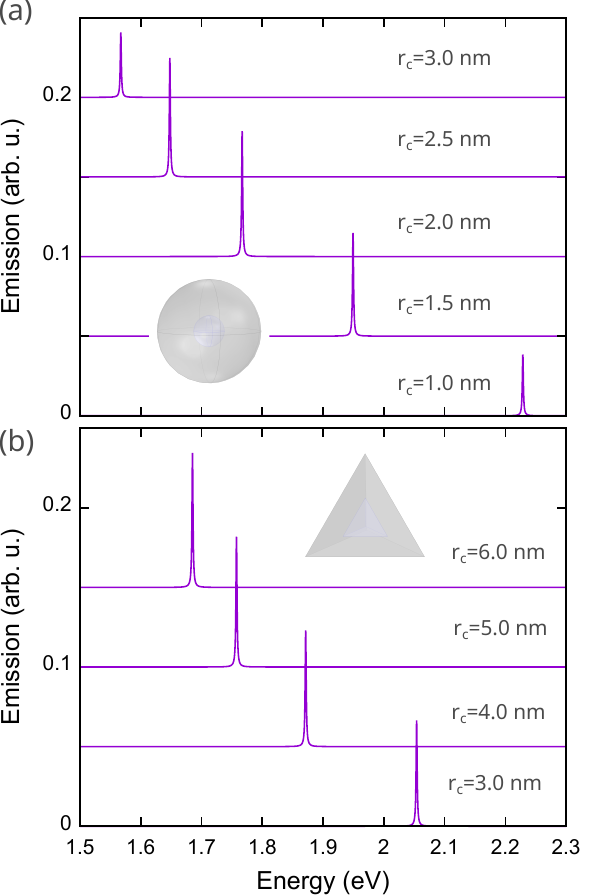}
	\caption{Emission spectrum of $X$ in InP/ZnSe QDs with (a) spherical and (b) tetrahedral shape, as a function of the core size. The population of states is calculated at $T=300$ K.  Shell sizes are the same as in Fig.~\ref{fig4}.
	}
    \label{fig5}
\end{figure}

The different nature of the hole ground state in large spherical or tetrahedral QDs 
manifests in the emission spectrum as well. 
In Figure \ref{fig5} we plot the calculated spectra at $T=300$ K.
For spherical QDs, Fig.~\ref{fig5}a, with increasing $r_c$ 
the band edge emission first gains oscillator strength
 --because the $1S_e$ state becomes more localized in the core--,
 but then decreases again (see $r_c \geq 2.5$ nm). 
 The latter effect is because the bright $1S_{3/2} 1S_e$ exciton
 starts sharing population with the dark $1P_{3/2} 1S_e$ one.
 No such effect is however observed in tetrahedral QDs, Fig.~\ref{fig5}b.

 The fact that deviations between spherical and tetrahedral shape reveal for large QDs,
 as shown in this section and in the previous one, is somewhat surprising. 
 One could expect them to show up in small QDs instead, when
 tetrahedral confinement is sensed more strongly.
 The underlying reason is that the cubic lattice symmetry
 is ultimately responsible for the $T_d$ features.
 In zinc-blende QDs, cubic band warping terms are often neglected,
 and the spherical approximation provides a good description of
 the energy spectrum. In the weak confinement regime, however,
 the level spacing becomes small enough that band warping
 terms mix the states and re-establish the underlying cubic
 symmetry of the lattice.

 \subsection{Charged excitons and biexcitons}

\begin{figure}[h]
    \centering
    \includegraphics[width=7.5cm]{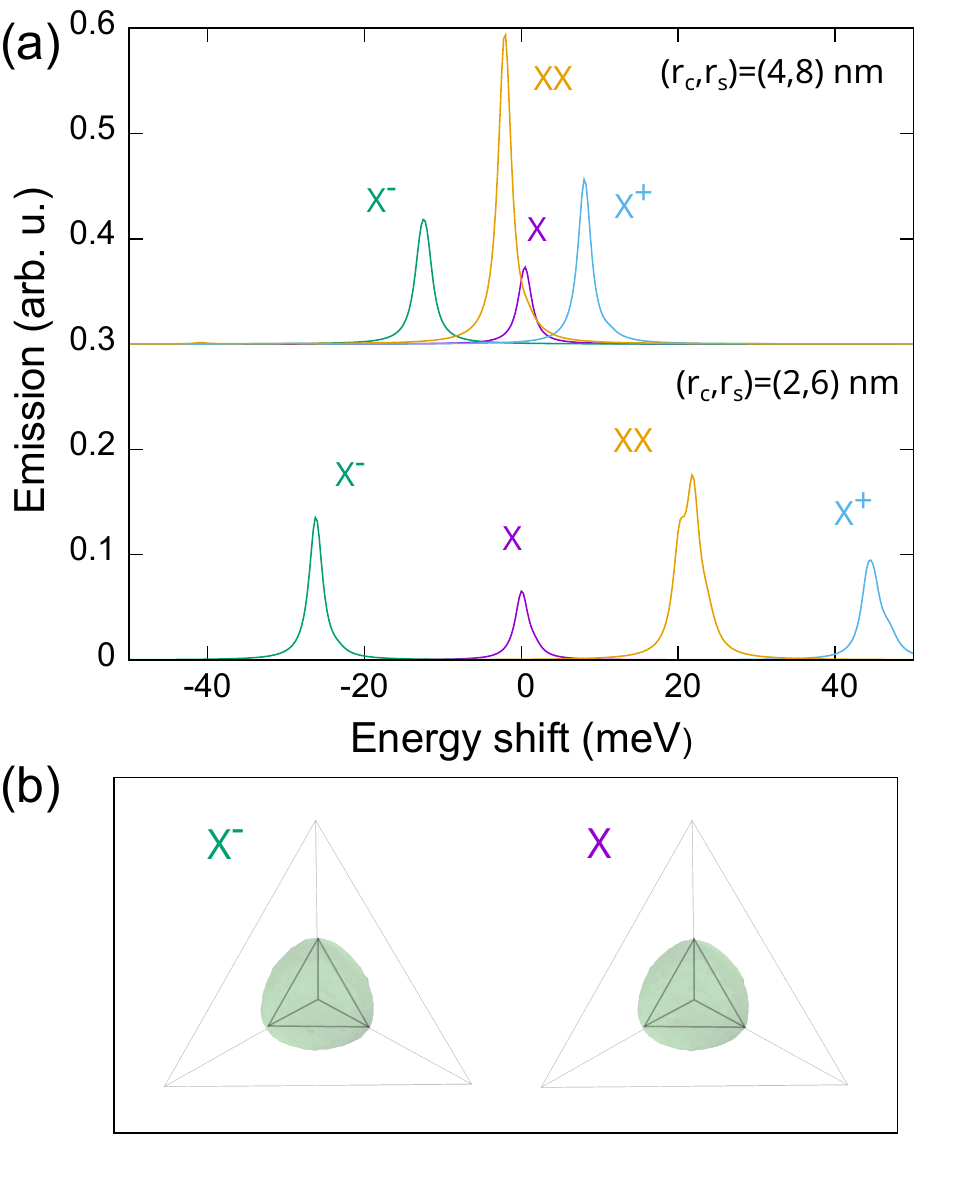}
	\caption{(a) Emission spectrum of excitons, trions and biexcitons in two InP/ZnSe QDs with different sizes.
    A temperature of $T=300$ K is assumed to calculate the population of states. The reference energy is that of $X$ emission.
	(b) Charge density of a single electron within the negative trion (left) and exciton (right) ground state,
	for the QD with $(r_c,r_s)=(2,6)$ nm.
	The isosurfaces contain 80\% of the charge density.}
    \label{fig6}
\end{figure}

The effect of multi-carrier interactions is studied in Figure \ref{fig6}.
In Fig.~\ref{fig6}a we compare the emission of excitons, trions and biexcitons in two tetrahedral QDs with different sizes.
One can see that $X^-$ is systematically redshifted with respect to the exciton (that is, it is bound),
while $X^+$ is systematically blueshifted (antibound). In turn, $XX$ can switch from antibound to bound depending on the QD dimensions.
For all the excitonic species and QD sizes we investigate, electron correlations are weak. 
The spectroscopic shifts observed in the figure thus arise mainly from first order Coulomb perturbations.
For example, in the QD with $(r_c,r_s)=(2,6)$ nm, the non-interacting ground state has 
$\langle V_{ee} \rangle = 249$ meV, $\langle V_{eh} \rangle=-273$ meV and $\langle V_{hh} \rangle = 331$ meV.
That is, electron repulsions are weaker because charges partially delocalize into the shell,
hole repulsions are stronger because of the almost complete localization into the core, and exciton attractions are in between.
 Consequently, in $X^-$ attractions prevail over attractions, which results in a redshift (binding energy).
 The first order perturbation estimate is $\Delta E_{X^-}^{(1)} = \langle V_{eh} + V_{ee} \rangle = -24$ meV,
 in close agreement with the CI calculation of Fig.~\ref{fig6}a. 
 On the other hand, in $X^+$ repulsions prevail over attractions.  The perturbational estimate, 
 $\Delta E_{X^+}^{(1)} = \langle V_{eh}+V_{hh} \rangle = 58$ meV is somewhat larger than the
blueshift of $\Delta E_{X^+} = 44$ meV obtained from CI calculations, 
which indicates that hole-hole correlations do relax repulsions. 

 The weakness of electronic correlations implies that the charge density of $X$ and $X^-$ is similar despite 
 electron-electron repulsions. This can be seen in Fig.~\ref{fig6}b, where we compare isosurfaces containing
 80\% of the electron charge density within $X^-$ (left) and $X$ (right). The delocalization into the shell is nearly identical,
 even though we have chosen a small core ($r_c=2$ nm) to promote tunneling.
 This result is at odds with the interpretation of Auger decay rates of $X^-$ in InP/ZnSe QDs, given in Ref.~\onlinecite{KimNL}.
 With increasing ZnSe shell thickness, the authors observed slower Auger rates, which was tentatively ascribed to 
 a greater delocalization of electrons into the shell, fostered by repulsions. 
 In view of our results, we posit the observed trend may rather be connected with the suppressed dielectric confinement
 (and hence weaker Coulomb interactions) in thick shell InP/ZnSe QDs.
  
 As mentioned before, the perturbative character of Coulomb interactions involving electrons, despite 
 the strong repulsions ($\langle V_{ee} \rangle \sim 250$ meV), is a consequence of the core/shell
 structure. As shown in Fig.~\ref{fig1}d, $1S_e$ electrons are split from $1P_e$ and $2S_e$ states by 
 $\Delta E \geq 300$ meV. That is, confinement energies exceed Coulomb ones. This is true for all the core sizes
 we study. %The reason is that the $1S_e$ orbital is significantly stabilized by the core, while 
% excited orbitals, with a node inside the core, are less so. As a result, even if $r_c$ increases,
% the separation between electron states does not decrease much.
 %The situation is different for holes, Fig.~\ref{fig1}(e,f), where the states pack together in an energy
% range under $V_{hh}$, so that correlations are more important. 

 A final comment is worth on the ground state degeneracy of the excitonic species. 
 In the tetrahedral QDs, disregarding electron-hole exchange interaction, we obtain 
 an 8-fold degeneracy for $X$, 4-fold for $X^-$, 10-fold for $X^+$ and 5-fold for $XX$.
 This result is consistent with that expected in spherical, zinc-blende QDs under the influence
 of valence band coupling.\cite{RodinaPRB}  In a non-interacting particle scheme,
 electrons have a total Bloch angular momentum $j=1/2$ and holes $j=3/2$. 
 $1S_e$ and $1S_{3/2}$ states are then 2-fold and 4-fold degenerate, respectively,
 which explains the 8-fold degeneracy of $X$. 
 Two interacting electrons form a singlet ground state with $J_e=0$, which explains
 the 4-fold degeneracy of $X^-$. 
 In turn, two interacting holes give rise to $J_h= 3/2\times 3/2=3 \oplus [2] \oplus 1 \oplus [0]$.
 Only $J_h=2$ and $J_h=0$ are antisymmetric with respect to permutation, and the former is slightly
 stabilized by hole-hole exchange interaction. This explains the 10-fold degeneracy of $X^+$
 and the 5-fold degeneracy of $XX$.
 The $J_h=0$ two-hole state is only a few meV away in energy, and it is responsible for the
 asymmetric bandshape of $XX$ and $X^+$ in Fig.~\ref{fig6}a (bottom spectrum).
 
% Recent experiments on shell-filling of InP/ZnSe QDs reported a 2-fold degeneracy for the conduction band,
% combined with a 5- to 9-fold degeneracy of the valence band.\cite{VelosaAOM}
% This deviates not only from single-band expectations, but also from the 4-fold degeneracy of $1S_{3/2}$.
% We argue that the high degeneracy may be the result of coupling holes with $j=3/2$,
% which gives ground states with non-trivial multiplicities. {\bf (???) % tmp ***
 %the presence of $X^+$ in the experiments may explain the observations.
% The formation of $X^+$ in InP/ZnSe QDs has been verified by means of magneto-luminescence.\cite{TolmachevACS}

\section{Conclusions}

By means of multi-band k$\cdot$p Hamiltonians, combined with CI simulations and 
symmetry point group analysis, we have shown that:\\

(i) The near-band-edge electronic structure of InP/ZnSe QDs with tetrahedral shape 
and cubic lattice is similar to that of spherical QDs in the quasicubic ($\mu=0$) lattice approximation, 
in terms of degeneracies and optical selection rules.
Deviations arise for large (dark red emitting) QDs only, where $S_{3/2}$ and $P_{3/2}$ states
(both with $\Gamma_8$ symmetry in the $\overline{T}_d$ group) admix.\\ 
 
(ii) Valence band mixing is key to understanding the symmetry, composition and degeneracy of hole states.
HH-LH coupling is as strong as in spherical, zinc-blende QDs\cite{EfrosPRB}, and SOH has non-neglegible
contributions, including the formation of low-energy $1S_{1/2}$ states.\\
 
(iii) Excitonic interactions are mostly perturbative for low-lying states, which reinforces the
spectral assignment (based on a non-interacting model) of Ref.~\onlinecite{RespektaACS}.\\

(iv) Electron-electron interactions are mostly perturbative as well.  
Repulsions in $X^-$ do not lead to a significant increase of the electron delocalization into the shell. 
The perturbative character of the Coulomb interactions involving electrons is because the $1S_e$ state
is in a strong confinement regime. Unlike excited states, its penetration into the ZnSe shell is
moderate. The localization in the core stabilizes the state, leading to energy gaps with $1P_e$ 
and $2S_e$ states exceeding Coulomb repulsions.

% against quasi-type-II model of Kelley and KimNL?

\begin{acknowledgments}
We acknowledge support from Grant No. PID2024-162489NB-I00, funded by Ministerio de Ciencia, Innovaci\'{o}n y Universidades 
(MICIU/AEI/10.13039/501100011033/ FEDER, EU).

The data that support the findings of this article are openly available.\cite{Dataset} 
\end{acknowledgments}

\appendix

\section{Character table and products of irreducible representations in $\overline{T}_d$}\label{ap:prods}

To facilitate the discussion, we reproduce here the character table of the $\overline{T}_d$ group\cite{Dresselhaus_book}
using Mulliken notation.
The dimension of the irreducible representation gives the degeneracy of the states (e.g. 4 for $\Gamma_8$). The character under $\bar{C}_3$ permits discriminating $\Gamma_7$ and $\Gamma_8$ states, as explained in Appendix \ref{ap:sim}.
The tables of products are used to determine the selection rules in Section \ref{ss:rules}.

\begin{table*}
	\centering
	\caption{Character table of the double group $\overline{T}_d$. 
	}\label{tab:chars}
\[
\begin{array}{|l|c|c|c|c|c|c|c|c|c|c|c|}
\hline
\text{Irrep} & E & 8C_3 & 3C_2 & 6S_4 & 6\sigma_d & \overline{E} & 8\overline{C}_3 & 3\overline{C}_2 & 6\overline{S}_4 & 6\overline{\sigma}_d & \text{Example of basis}\\
\hline
\Gamma_1 & 1 & 1 & 1 & 1 & 1 & 1 & 1 & 1 & 1 & 1 & s \; \text{(spherical orbital)} \\
\hline
	\Gamma_2 & 1 & 1 & 1 & -1 & -1 & 1 & 1 & 1 & -1 & -1 & \text{pseudoscalar} \\
\hline
\Gamma_3 & 2 & -1 & 2 & 0 & 0 & 2 & -1 & 2 & 0 & 0 & (2z^2-x^2-y^2, x^2-y^2) \\
\hline
	\Gamma_4 & 3 & 0 & -1 & 1 & -1 & 3 & 0 & -1 & 1 & -1 & \text{Rotations} \; (R_x,R_y,R_z) \\
\hline
\Gamma_5 & 3 & 0 & -1 & -1 & 1 & 3 & 0 & -1 & -1 & 1 & (x,y,z); (xy,yz,zx) \\
\hline
\Gamma_6 & 2 & 1 & 0 & \sqrt{2} & 0 & -2 & -1 & 0 & -\sqrt{2} & 0 & (u_1, u_2) \\
\hline
\Gamma_7 & 2 & 1 & 0 & -\sqrt{2} & 0 & -2 & -1 & 0 & \sqrt{2} & 0 & (u_7, u_8) \\
\hline
\Gamma_8 & 4 & -1 & 0 & 0 & 0 & -4 & 1 & 0 & 0 & 0 & (u_3, u_4, u_5, u_6)  \\
\hline
\end{array}
\]
\end{table*}

\begin{table*}
	\centering
	\caption{Product of irreducible representations in the double group $\overline{T}_d$.}\label{tab:prods}
\[
\begin{array}{|c|c|c|c|c|c|c|c|c|}
\hline
\otimes & \Gamma_1 & \Gamma_2 & \Gamma_3 & \Gamma_4 & \Gamma_5 & \Gamma_6 & \Gamma_7 & \Gamma_8 \\
\hline
\Gamma_1 & \Gamma_1 & \Gamma_2 & \Gamma_3 & \Gamma_4 & \Gamma_5 & \Gamma_6 & \Gamma_7 & \Gamma_8 \\
\hline
\Gamma_2 &  & \Gamma_1 & \Gamma_3 & \Gamma_5 & \Gamma_4 & \Gamma_7 & \Gamma_6 & \Gamma_8 \\
\hline
\Gamma_3 &  &  & \Gamma_1 \oplus \Gamma_2 \oplus \Gamma_3 & \Gamma_4 \oplus \Gamma_5 & \Gamma_4 \oplus \Gamma_5 & \Gamma_8 & \Gamma_8 & \Gamma_6 \oplus \Gamma_7 \oplus \Gamma_8 \\
\hline
\Gamma_4 &  &  &  & \Gamma_1 \oplus \Gamma_3 \oplus \Gamma_4 \oplus \Gamma_5 & \Gamma_2 \oplus \Gamma_3 \oplus \Gamma_4 \oplus \Gamma_5 & \Gamma_6 \oplus \Gamma_8 & \Gamma_7 \oplus \Gamma_8 & \Gamma_6 \oplus \Gamma_7 \oplus 2\Gamma_8 \\
\hline
\Gamma_5 &  &  &  &  & \Gamma_1 \oplus \Gamma_3 \oplus \Gamma_4 \oplus \Gamma_5 & \Gamma_7 \oplus \Gamma_8 & \Gamma_6 \oplus \Gamma_8 & \Gamma_6 \oplus \Gamma_7 \oplus 2\Gamma_8 \\
\hline
\Gamma_6 &  &  &  &  &  & \Gamma_1 \oplus \Gamma_4  & \Gamma_2 \oplus \Gamma_5 & \Gamma_3 \oplus \Gamma_4\oplus \Gamma_5 \\
\hline
\Gamma_7 &  &  &  &  &  &  & \Gamma_1 \oplus \Gamma_4 & \Gamma_3 \oplus \Gamma_4 \oplus \Gamma_5 \\
\hline
\Gamma_8 &  &  &  &  &  &  &  & \Gamma_1 \oplus \Gamma_2 \oplus \Gamma_3 \\
&  &  &  &  &  &  &  & \oplus 2\Gamma_4 \oplus 2\Gamma_5 \\
\hline
\end{array}
\]
\end{table*}

\section{Determination of the symmetry of states in $\overline{T}_d$}\label{ap:sim}

The identification of $\Gamma_8$ states is straightforward because they are four-fold degenerate 
(see Table \ref{tab:chars}). To discriminate the $\Gamma_6$ and $\Gamma_7$ doublets, 
we notice that they have different characters under the $C_3$: $\chi^{(\Gamma_6)}(C_3)=1$ and $\chi^{(\Gamma_7)}(C_3)=-1$.
We label the states forming the doublet $F_1$ and $F_2$.
 Visual identification of the symmetry under $C_3$ is obscured by the degeneracy of $F_1$ and $F_2$. 
 The character can however be calculated as:
 \begin{multline}
 \chi(C_3) = \sum_{j=1}^2 \langle F_j \mid C_3 \mid F_j \rangle = \\
	 \sum_{j=1}^2 \sum_{k} \lambda_k \int d^3r \; 
	 f_{j,k}^*(\mathbf r)\, 
	 f_{j,k}(C_3^{-1}\mathbf r).
 \end{multline}
 \noindent In this expression, $k=1,2$ for electron states,
 while $k=3,8$ for hole states.  $\lambda_k$ is the eigenvalue
 of $C_3$ when acting on the Bloch function $|u_k\rangle$.
 In general, $C_3\,|j,m_j\rangle = e^{i m_j 2\pi/3}\,|j,m_j\rangle$.
 For the Bloch functions in Eqs.~(\ref{eq:elec},\ref{eq:HHLH},\ref{eq:SOH}),
 the following relations emerge:
	 $\lambda_1= e^{+i\pi/3}$,
	 $\lambda_2= e^{-i\pi/3}$,
	 $\lambda_3= -1$,
	 $\lambda_4= e^{+i\pi/3}$,
	 $\lambda_5= e^{-i\pi/3}$,
	 $\lambda_6= -1$,
	 $\lambda_7= e^{+i\pi/3}$,
	 $\lambda_8= e^{-i\pi/3}$.

 %
% \begin{eqnarray}
%	 \nonumber
%	 |u_1\rangle &\rightarrow& \lambda_1= e^{+i\pi/3}, \\
%	 \nonumber
%	 |u_2\rangle &\rightarrow& \lambda_2= e^{-i\pi/3}, \\
%	 \nonumber
%	 |u_3\rangle &\rightarrow& \lambda_3= -1, \\
%	 \nonumber
%	 |u_4\rangle &\rightarrow& \lambda_4= e^{+i\pi/3}, \\
%	 \nonumber
%	 |u_5\rangle &\rightarrow& \lambda_5= e^{-i\pi/3}, \\
%	 \nonumber
%	 |u_6\rangle &\rightarrow& \lambda_6= -1, \\
%	 \nonumber
%	 |u_7\rangle &\rightarrow& \lambda_7= e^{+i\pi/3}, \\
%	 \nonumber
%	 |u_8\rangle &\rightarrow& \lambda_8= e^{-i\pi/3}.
% \end{eqnarray}

\bibliography{InP_biblio}

\end{document}